# Topological magnon noises

Shuang Liang[1†], Tengyue Zhao[1†], Yu-Hang Li[1*], and Hua Jiang[2]

**Affiliations:**

1 Department of Physics, Nankai University, Tianjin 300071, China

2 Interdisciplinary Center for Theoretical Physics and Information Sciences, Fudan University, Shanghai 200433, China

*Corresponding author. Email: liyuhang@nankai.edu.cn

†These authors contributed equally to this work.

**Abstract:** We develop a comprehensive formalism for magnon transport in ferromagnetic insulators driven by a temperature gradient. The formulas for magnon current and corresponding magnon noise are derived herein based on the spin Hamiltonian of a topological magnon insulator, which enables us to calculate the magnon Hall angle, to provide an explicit expression for the Fano factor, and to reaffirm the quantitative relations between magnon conductance and magnon noise. We find that the magnon current is not conserved in the presence of the Gilbert damping. Consequently, the reciprocal relation between the local and nonlocal noises, the Johnson-Nyquist formula between the conductance and the thermal noise, and the relation between the transmission coefficient and the shot noise are profoundly altered.

**Introduction:** Transport properties are fundamental building blocks of condensed matter physics that has been well established as a powerful tool in exploring novel physics and applications [1-4]. They mainly consist of two ingredients: one dubbed current concerning to the exchange of physical observables such as charge, energy, or momentum, and the other corresponds to the fluctuation and correlation of this current known as current noise [5-8], including the thermal noise in equilibrium and the nonequilibrium shot noise. While current provides the time averaging transport information, noise reveals in-depth insight into the non-equilibrium dynamics. For example, a quantum anomalous Hall insulator exhibits a quantized conductance $e^2/h$ in a two terminal system ascribed to the chiral edge state [9-11], which can be alternatively characterized by the noise signature [12].

In principle, such a system can be intuitively understood by using a single channel scattering model with a quantized transmission probability $\sigma = 1$[5,7]. Due to the charge conservation, the current flowing into the central quantum anomalous Hall insulator and that flowing out are opposite to each other but equal in magnitude. Therefore, the local current noise in one side is by definition reciprocal to the non-local one across the central tunnel barrier. At zero temperature, the thermal noise vanishes because of the blockage of thermal agitation while the shot noise can be explicitly expressed as $\mathcal{S} = 2e^3V\sigma(1-\sigma)/h$ where $e$ is the electron charge, $h$ is the Planck constant and $V$ is the voltage drop across the system. By contrast, at finite temperature, the thermal noise in equilibrium reproduces the Johnson-Nyquist formula $\mathcal{T} = 4Gk_BT$ with $G$ the system conductance, $k_B$ the Boltzmann constant and $T$ the temperature, overwhelming the shot noise $\mathcal{S}$.

The quantum anomalous Hall effect has been extended to magnon systems where a nontrivial magnon band emerges [13-17], resulting in a magnon Hall effect supporting a similar chiral magnon edge mode [18,19]. A natural question is whether and how the noise signature manifests in such a topological magnon insulator (TMI). Different from electron, magnon is boson, whose transport is usually driven by a statistical force such as the temperature gradient and thus cannot be understood from the heuristic scattering model [20,21]. Moreover, the differences between magnon and electron also bring in some contradictions to the quantum anomalous Hall effect. First, magnon can only be populated by the thermal energy at finite temperature and obeys the Bose-Einstein statistics rather than Fermi-Dirac distribution. Second, because of some dissipation effect such as the Gilbert damping or magnetic disorders, magnon number is not conserved. These contradictions significantly enhance the fluctuations of magnon during transport process, making its noise properties more essential than that in electronic systems. On the other hand, it has been shown that both the edge magnon mode and the bulk magnon bands contribute to the magnon current including the transverse magnon Hall current if the Berry curvature is non-vanishing [18,22], indicating that the magnon noises are intractable to treat suitably.

In this work, we resolve these puzzles by showing how noise signature manifests in a TMI. We overcome the aforementioned issues by utilizing the non-equilibrium Green's function method and derive the general formalism for temperature gradient driven magnon current and corresponding magnon noise. It turns out that both the magnon current and magnon Hall current are not conserved in the presence of the dissipation effect. Consequently, the reciprocal relation between the local and nonlocal magnon noises is not maintained. Moreover, it is found that although the magnon transmission coefficient and the kernel of the magnon noise are still implicitly relevant, the Johnson-Nyquist relation between the magnon conductance and the thermal magnon noise, the relation between magnon transmission coefficient and the shot magnon noise do not hold. We also calculate the Fano factor and the magnon Hall angle in order to gain important insights into the transport properties of a TMI.

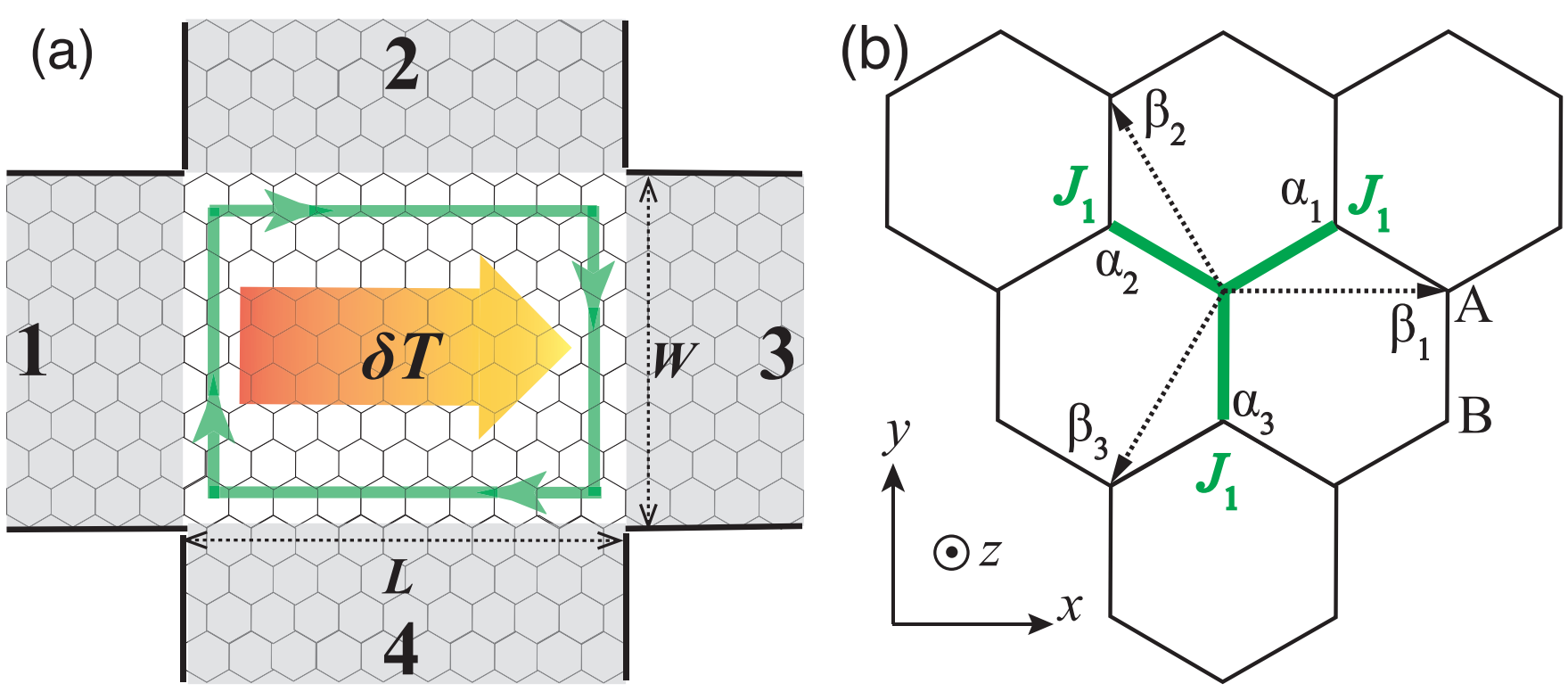

Fig. 1 (a) Schematic of the four terminal TMI transport system. The system size is $L \times W$ along $x$ and $y$ directions, respectively. A temperature gradient $\delta T$ is applied along $x$-direction, which drives a bulk magnon transport represented by a wide gold arrow and simultaneously a chiral edge magnon transport denoted by the green line. (b) The TMI on a honeycomb ferromagnetic insulator. The nearest and next nearest neighbor bonds are labeled by $\alpha_i$ and $\beta_i$, respectively.

The magnon transport through the central ferromagnetic insulator in a multiterminal system, illustrated in Fig. 1(a), can be simulated by the Hamiltonian

$$H = H_C + \sum_m H_m + H_t. \quad (1)$$

Here, $H_C = (3JS + 2K)\sum_i c_i^\dagger c_i - JS\sum_{\langle ij\rangle}(c_i^\dagger c_j + H.c.) - DS\sum_{\langle\langle ij\rangle\rangle}(i\epsilon_{ij} c_i^\dagger c_j + H.c.)$ is the Hamiltonian for the central TMI with spin-$S$ [15,16,23], where $J > 0$ is the nearest neighbor exchange interaction, $K > 0$ is the uniaxial anisotropy, $D$ is the next nearest neighbor Dzyaloshinskii-Moriya interaction with $\epsilon_{ij} = \pm 1$ depending on the chirality and $c_i^\dagger$ ($c_i$) creates (annihilates) a magnon at site $i$. The Hamiltonian $H_C$ can be derived from the spin Hamiltonian on a honeycomb lattice as schematically shown in Fig. 1(b) by performing the Holstein-Primakoff transformation [23]. $H_m$ is the magnon Hamiltonian for lead *m*. $H_t$ describes the coupling between the scattering region and all leads. Different temperature $T_m$ and $T_C$ are assigned to different leads and the central region, respectively. Some dissipation effect such as the Gilbert damping or magnetic disorders is allowed in the central scattering region.

The magnon current flowing through the interface between lead *m* and the central scattering region can be calculated from the time evolution of the magnon number operator in that lead $J_m = -\sum_{j\in m}\langle \partial_t \hat{N}_j\rangle = i/\hbar \sum_{j\in m}\langle [\hat{N}_j,\ H]\rangle$, where $\hat{N}_j$ is the magnon number operator and $\langle\cdots\rangle$ denotes the ensemble average of the physical observable. Applying the commutation relation recasts the magnon current $J_m$ as

$$J_m = \frac{1}{\hbar}\mathrm{Tr}[\boldsymbol{T}_{mC}\boldsymbol{G}_{Cm}^{<}(t,t) + \mathrm{H.c.}], \quad (2)$$

where $\boldsymbol{T}_{mC} = \boldsymbol{T}_{Cm}^\dagger$ refers to the coupling matrix from lead *m* to the central TMI in the Hamiltonian $H_t$ and the cross Green's function $\boldsymbol{G}_{Cm}^{<}(t,t)$ is defined as $\boldsymbol{G}_{Cm}^{<}(t,t') \equiv -i\sum_{j'\in m, j\in C}\langle a_{j'}^\dagger(t')c_j(t)\rangle$. In accordance with the same procedure of the non-equilibrium Green's function method for phonon transport [24], equation 2 can be straightforwardly rewritten as

$$J_m = \frac{1}{\hbar}\int_0^\infty d\epsilon \mathrm{Tr}[\boldsymbol{\Sigma}_m^{>}(\epsilon)\boldsymbol{G}_C^{<}(\epsilon) - \boldsymbol{\Sigma}_m^{<}(\epsilon)\boldsymbol{G}_C^{>}(\epsilon)]. \quad (3)$$

Here, $\boldsymbol{G}_C^{<(>)}(\epsilon)$ is the Green's function of the central TMI. The self energy due to coupling to the lead $m$ is $\boldsymbol{\Sigma}_m^{<(>)}(\epsilon) = -i[f(\epsilon) + 1/2 \mp 1/2]\boldsymbol{\Gamma}_m(\epsilon)$, where $f(\epsilon) = 1/[e^{\frac{\epsilon}{k_B T_m}} - 1]$ is the Bose-Einstein distribution function and $\boldsymbol{\Gamma}_m(\epsilon) = i[\boldsymbol{\Sigma}_m^r(\epsilon) - \boldsymbol{\Sigma}_m^a(\epsilon)]$ is the line width function, which can be obtained through the mode matching method [25]. Using the Keldysh equation $\boldsymbol{G}_C^{<(>)}(\epsilon) = \boldsymbol{G}_C^r(\epsilon)\sum_n \boldsymbol{\Sigma}_n^{<(>)}(\epsilon)\,\boldsymbol{G}_C^a(\epsilon)$ and the Dyson equation $\boldsymbol{G}_C^{r(a)}(\epsilon) = \left[\epsilon \pm i\gamma - H_C - \sum_n \boldsymbol{\Sigma}_n^{r(a)}(\epsilon)\right]^{-1}$ with $\gamma$ the Gilbert damping, we obtain

$$J_m = \frac{1}{\hbar}\int_0^\infty d\epsilon[f_m(\epsilon) - f_n(\epsilon)]\sigma_{mn}(\epsilon), \quad (4)$$

where the magnon transmission coefficient from lead *n* to lead *m* reads

$$\sigma_{mn}(\epsilon) = \mathrm{Tr}[\boldsymbol{\Gamma}_n(\epsilon)\boldsymbol{G}_C^r(\epsilon)\boldsymbol{\Gamma}_m(\epsilon)\boldsymbol{G}_C^a(\epsilon)]. \quad (5)$$

Equation 4 is the general form for temperature gradient driven magnon current of a multi terminal ferromagnetic insulator. A similar expression for a two terminal case has been given in 13. In the limit of small temperature gradient, the magnon current in Eq.4 can be further expressed as

$$J_m = \sum_{m,n} \frac{T_n - T_m}{\hbar T} \int_0^\infty d\epsilon \epsilon \partial_\epsilon f_T(\epsilon) \sigma_{mn}(\epsilon), \tag{6}$$

where $T = T_C$ is the system temperature. We stress that this expression is different from that derived in 20 and 21. Since Eqs. 4 and 6 are derived from the spin Hamiltonian in real space, we can conveniently incorporate the magnon dissipation effect such as the Gilbert damping or magnetic disorders. Moreover, they can be used to calculate not only the magnon Hall current but also the longitudinal one, and even the general magnon current in other trivial magnon insulators.

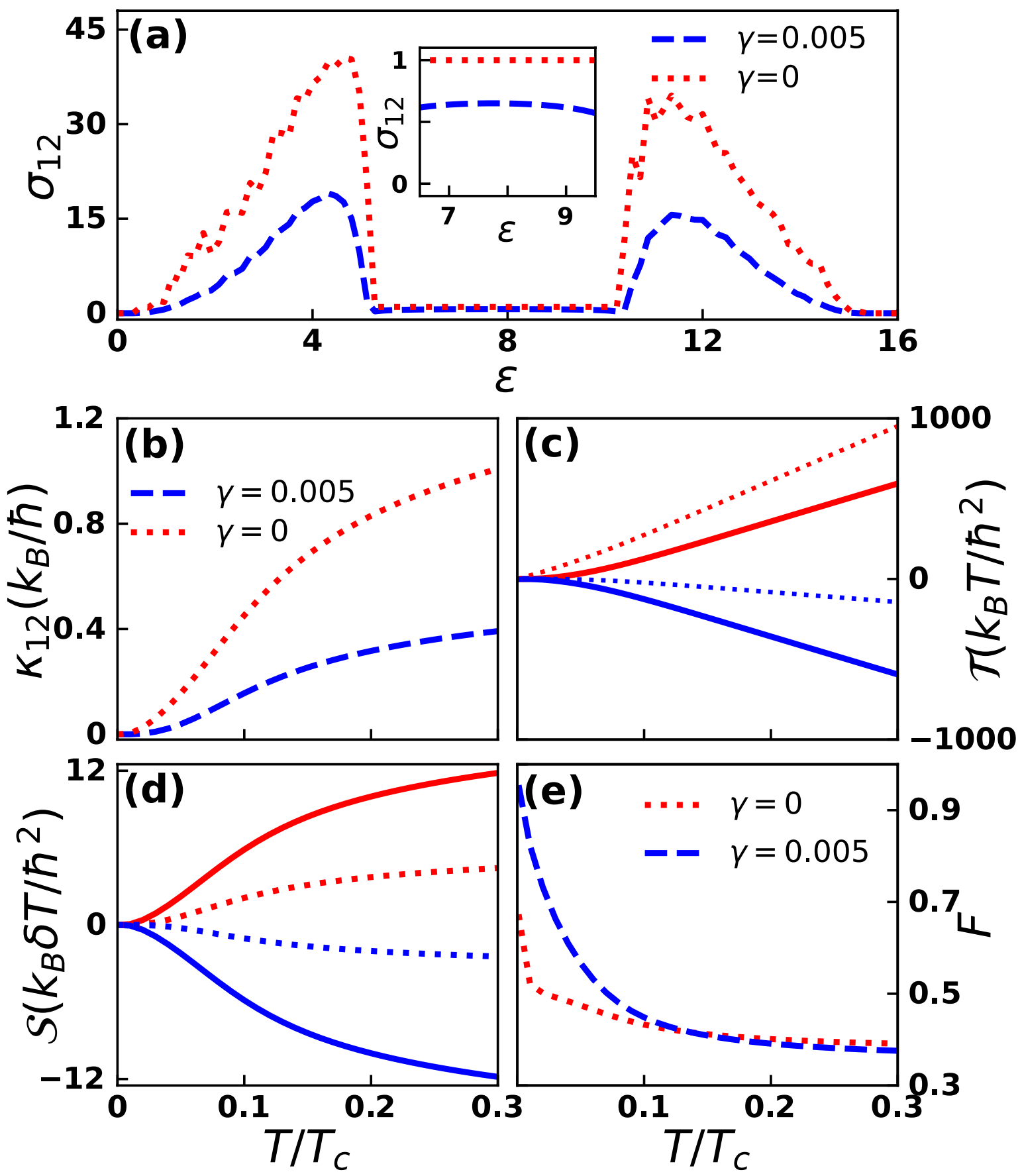

Fig. 2: (a) magnon transmission coefficients for a two-terminal system as a function of energy with (solid lines) and without (dashed lines) Gilbert damping. (b) The corresponding two terminal magnon conductance as a function of temperature. Local (red lines) and nonlocal (blue lines) thermal (c) and shot (d) magnon noises as a function of temperature with (solid lines) and without (dashed lines) Gilbert damping. (e) Temperature dependence of the Fano factor. Here, the temperature is scaled with respect to the Curie temperature $T_c = JS(S+1)/k_B$.

Hereinafter, we move on to the magnon noise defined as [5]

$$S_{mn}(\omega) = \int d\tau e^{i\omega\tau} \langle \delta\hat{J}_m(t) \delta\hat{J}_n(t+\tau) \rangle, \tag{7}$$

where $\delta\hat{J}_m(t) = \hat{J}_m(t) - \langle \hat{J}_m(t) \rangle$ with $\hat{J}_m(t)$ the magnon current operator. In steady state, the current operator is time independent. Therefore, it is convenient to work in the frequency space. In particular, we consider here the zero-frequency magnon noises $S_{mn}(\omega = 0)$, which reflects the magnon current fluctuations and is given by [5,7]

$$S_{mn}(\omega=0)=\frac{1}{\hbar^2}\int d\epsilon\kappa_{mn}(\epsilon), \tag{8}$$

where the kernel is [26]

$$\begin{aligned}\kappa_{mn}(\epsilon)=Tr[&\boldsymbol{G}_{Cm}^{<}(\epsilon)\boldsymbol{T}_{mC}(\epsilon)\boldsymbol{G}_{Cn}^{>}(\epsilon)\boldsymbol{T}_{nC}(\epsilon)\\&+\boldsymbol{T}_{Cn}(\epsilon)\boldsymbol{G}_{nC}^{<}(\epsilon)\boldsymbol{T}_{Cm}(\epsilon)\boldsymbol{G}_{mC}^{>}(\epsilon)\\&-\boldsymbol{G}_{C}^{<}(\epsilon)\boldsymbol{T}_{Cm}(\epsilon)\boldsymbol{G}_{mn}^{>}(\epsilon)\boldsymbol{T}_{nC}(\epsilon)\\&-\boldsymbol{T}_{Cn}(\epsilon)\boldsymbol{G}_{nm}^{<}(\epsilon)\boldsymbol{T}_{mC}(\epsilon)\boldsymbol{G}_{C}^{>}(\epsilon)].\end{aligned} \tag{9}$$

Using the Keldysh equation for the cross Green's function $\boldsymbol{G}_{mn}^{<(>)}(\epsilon)=\sum_p\boldsymbol{G}_{mp}^{r}(\epsilon)\left[f_p(\epsilon)+1/2\mp 1/2\right]\boldsymbol{\zeta}_p(\epsilon)\boldsymbol{G}_{pn}^{a}(\epsilon)$ with $\boldsymbol{\zeta}_p(\epsilon)=\left[\boldsymbol{g}_p^a(\epsilon)\right]^{-1}-\left[\boldsymbol{g}_p^r(\epsilon)\right]^{-1}$, the magnon noise in Eq. 8 finally becomes

$$S_{mn}(\omega=0)=\sum_{p,q}\frac{1}{\hbar^2}\int d\epsilon f_p(\epsilon)[1+f_q(\epsilon)]\pi_{mn}^{pq}(\epsilon), \tag{10}$$

where

$$\begin{aligned}\pi_{mn}^{pq}(\epsilon)=Tr[&\boldsymbol{G}_{Cp}^{r}\boldsymbol{\zeta}_p\boldsymbol{G}_{pm}^{a}\boldsymbol{T}_{mC}\boldsymbol{G}_{Cq}^{r}\boldsymbol{\zeta}_q\boldsymbol{G}_{qn}^{a}\boldsymbol{T}_{nC}\\&\boldsymbol{T}_{Cn}\boldsymbol{G}_{np}^{r}\boldsymbol{\zeta}_p\boldsymbol{G}_{pC}^{a}\boldsymbol{T}_{Cm}\boldsymbol{G}_{mq}^{r}\boldsymbol{\zeta}_q\boldsymbol{G}_{qC}^{a}\\&+i\boldsymbol{G}_{C}^{r}\boldsymbol{\Gamma}_p\boldsymbol{G}_{C}^{a}\boldsymbol{T}_{Cm}\boldsymbol{G}_{mq}^{r}\boldsymbol{\zeta}_q\boldsymbol{G}_{qn}^{a}\boldsymbol{T}_{nC}\\&+i\boldsymbol{T}_{Cn}\boldsymbol{G}_{np}^{r}\boldsymbol{\zeta}_p\boldsymbol{G}_{pm}^{a}\boldsymbol{T}_{mC}\boldsymbol{G}_{C}^{r}\boldsymbol{\Gamma}_q\boldsymbol{G}_{C}^{a}].\end{aligned} \tag{11}$$

Here in Eq. 11, we have omitted the variable $\epsilon$ for convenience. Equation 10 is the general form for the total noise signature including the shot magnon noise $\mathcal{S}_{mn}$ and the thermal magnon noise $\mathcal{T}_{mn}$. In the presence of a small temperature gradient, we can further separate them into

$$\mathcal{T}_{mn}=\sum_{p,q}\frac{-k_BT}{\hbar^2}\int d\epsilon\partial_\epsilon f_T(\epsilon)\pi_{mn}^{pq}(\epsilon), \tag{12-A}$$

$$\mathcal{S}_{mn}=\sum_{p,q}\frac{T_q-T_p}{2\hbar^2 T}\int d\epsilon\epsilon\partial_\epsilon f_T(\epsilon)\pi_{mn}^{pq}(\epsilon). \tag{12-B}$$

Equation 12 is the central result of this Letter. It is obvious that the shot magnon noise is considerably small compared to the thermal magnon noise. Furthermore, the thermal noise in this case possesses the general form $\mathcal{T}=4k_BT\mathcal{B}$ with $\mathcal{B}=-\sum_{p,q}\int d\epsilon\partial_\epsilon f_T(\epsilon)\pi_{mn}^{pq}(\epsilon)/4\hbar^2$ the effective magnon band width, restoring the conventional Johnson-Nyquist relation. In the absence of Gilbert damping$\gamma=0^+$, the kernels in Eq. 11 is related to the magnon conductance via the internal relation $\sum_{p,q}\pi_{mn}^{pq}(\epsilon)=\sigma_{mn}(\epsilon)$ in a two-terminal system. While inside the magnon band gap, only the chiral magnon edge mode contributes to the transport, we find that $\pi_{mn}^{p\neq q}(\epsilon)=\sigma_{mn}(1-\sigma_{mn})$, which further demonstrate the consistency and reliability of our theory. Nevertheless, different from that in quantum anomalous Hall insulators [12], the inclusion of the bulk magnon band violates the relation connecting the shot noise and the conductance.

Let us consider a TMI with parameters $S=5/2$, $J=1$, $K=0.05$, and $D=0.2$. The system size is taken as $L=50\sqrt{3}a_0$, $W=200a_0$ along $x$ and $y$ directions with $a_0=1$ the distance between two nearest neighbors. We assume that the magnon transport is driven by the temperature gradient along $x$- direction, and thus the temperature difference $\delta T = T_1 - T_3$ while $T_2 = T_4 = T$.

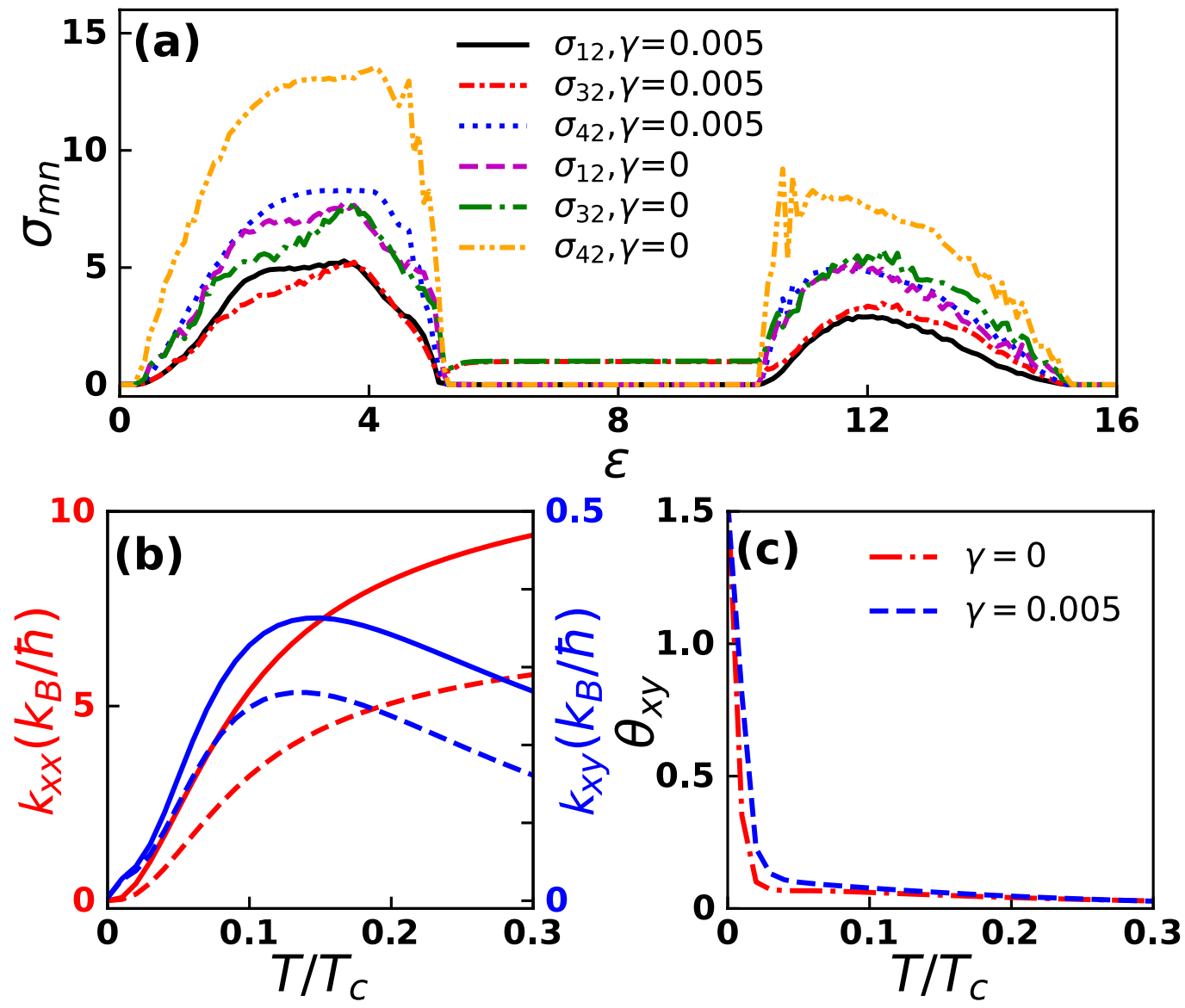

Fig. 3: (a) Energy dependence of some representative magnon transmission coefficients in a four-terminal system with and without Gilbert damping. (b) The longitudinal magnon conductance $\kappa_{xx}$ and the magnon Hall conductance $\kappa_{xy}$ as a function of temperature at $\gamma = 0$ (solid lines) and $\gamma = 0.005$ (dashed lines). (c) The corresponding magnon Hall angle. Here $\alpha = 0.5$.

In comparison to a quantum anomalous Hall insulator [12], we first consider a two-terminal case with the central scattering region connected to leads 1 and 3 only. In Fig. 2(a), we show the energy dependence of the magnon transmission coefficient $\sigma_{13}$ at $\gamma = 0$ and $\gamma = 0.005$. It is clear that the transmission coefficient quantizes identically to $\sigma_{13} = 1$ inside the magnon band gap when $\gamma = 0$, in good agreement with the quantized chiral magnon edge mode [16,19], which confirms the validity of our theory. Whereas the transmission coefficient is strongly suppressed in the presence of the Gilbert damping ($\gamma = 0.005$) because of the magnon dissipation induced by the Gilbert damping effect. Integrating the magnon transmission coefficient together with a prefactor $\epsilon \partial_\epsilon f_T(\epsilon)/\hbar T$ over the energy space by using Eq. 6 gives the two terminal magnon conductance $\kappa_{13} \equiv J_{13}/\delta T$, which is shown in Fig. 2(b) as a function of temperature for different Gilbert damping $\gamma$. We see that the magnon conductance $\kappa_{13}$ at $\gamma = 0.005$ is manifestly suppressed compared to that at $\gamma = 0$, both of which increase monotonically as the temperature is raised because the magnon density of states also increases with temperature.

It is worth noting that the magnon conductance shown here with the Gilbert damping at $\gamma = 0.005$ includes only the magnon current transport across the system. It excludes the contribution from the dissipation magnon current inside the central scattering region, which equals the reduction between the current flowing out from lead 1 and that flowing into lead 3. Such dissipation current breaks the magnon current conservation and can give rise to unbalanced local and nonlocal magnon noises. Indeed, as depicted in Figs. 2(c) and (d), both the local thermal magnon noise [Fig. 2(c)] and the local shot magnon noise [Fig. 2(d)] are larger than the nonlocal ones in the presence of the Gilbert damping.

To gain more insights into the magnon transport, in Fig. 2(e), we show the temperature dependence of the Fano factor defined as $F \equiv \mathcal{S}/2J$, which is a dimensionless factor that can be used to quantify the magnon noise. Unexpectedly, the Fano factor $F$ tends to 1 when the temperature approaches zero at $\gamma = 0.005$. This indicates that the magnon transport is discretized and that magnons are governed by the Poisson distribution in the limit $T \to 0$, which can only be transmitted one by one. As a result, the Fano factor predicted here provides a possible signature of single-magnon transport statistics [27,28].

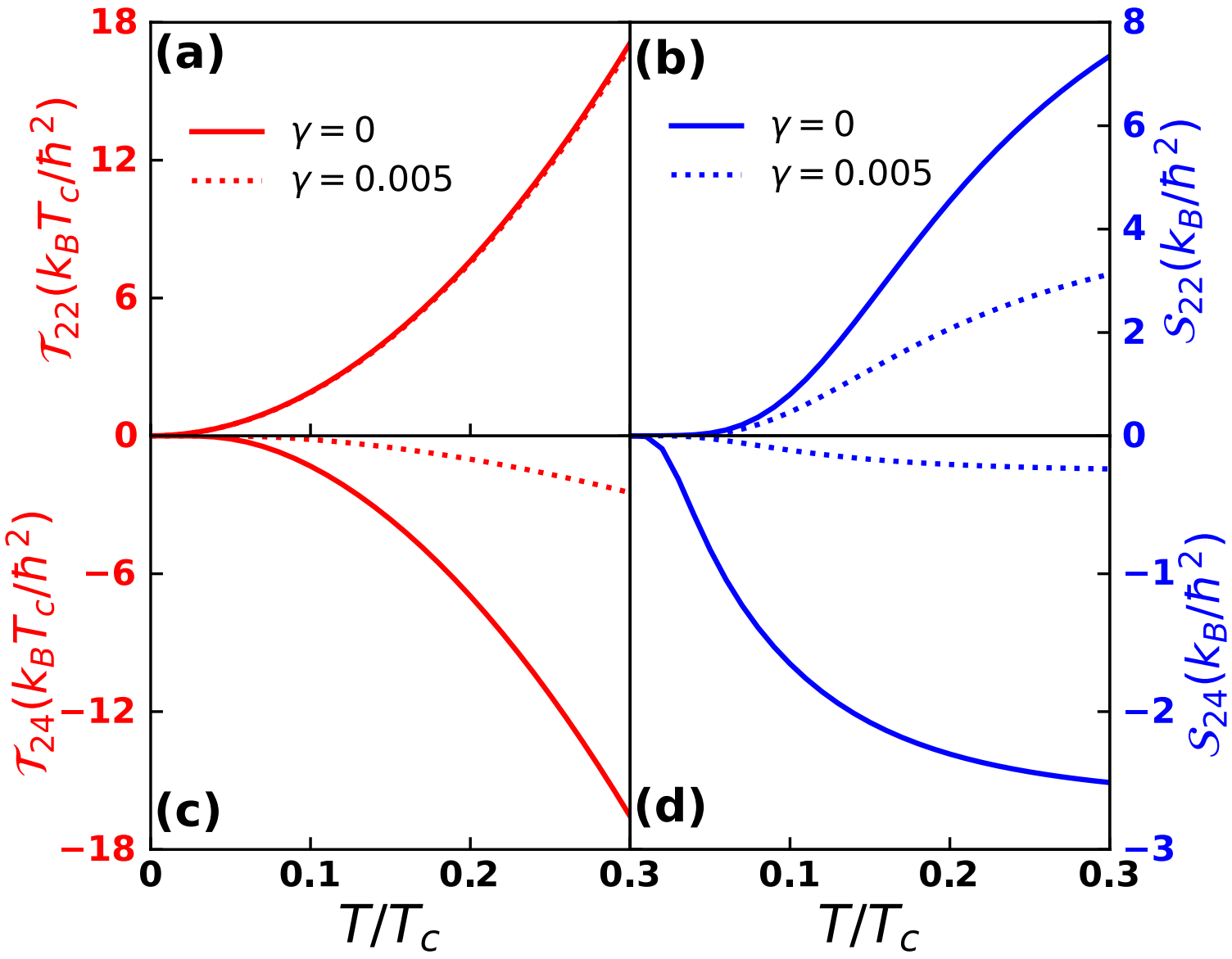

Fig. 4: Temperature dependence of the magnon noises in a four-terminal system. The left panel shows the local (a) and nonlocal (c) thermal magnon noises at $\gamma = 0$ and $\gamma = 0.005$ while the right panel shows the corresponding shot magnon noises. Here $\alpha = 0.5$.

To incorporate the magnon Hall transport, we then consider a four-terminal case as illustrated in Fig. 1(a). The temperature gradient is sustained along $x$-direction, and we assume that $T_1 = T + \alpha\delta T = T_3 + \delta T$ where $\alpha$ characterizes the temperature drop between lead 1 and the central scattering region. In this regard, the longitudinal magnon conductance $\kappa_{xx}$ and the magnon Hall conductance $\kappa_{xy}$ can still be obtained by using Eq. 6. In Fig. 3(a), we plot the magnon transmission coefficients from other leads to lead 2 as a function of energy $\epsilon$ in the presence (absence) of Gilbert damping $\gamma$. It shows that the transmission coefficient $\sigma_{32}$ is identical to 1 while $\sigma_{12}$ is zero inside the magnon band gap, affirming the novel magnon edge mode that can only propagate counterclockwise. Besides, we also see that even though both the bulk band and the edge mode can contribute to the magnon Hall transport, the magnon edge mode is more robust against the Gilbert damping $\gamma$, which agrees with the existing works [18,19].

Figure 3(b) shows the temperature dependence of the longitudinal magnon conductance $\kappa_{xx}$ and the magnon Hall conductance $\kappa_{xy}$ with and without the Gilbert damping. While $\kappa_{xx}$ increases monotonically with temperature, $\kappa_{xy}$ decreases after a critical temperature $T \approx 0.13T_c$, exhibiting a local maximum originates probably from the opposite chiralities of the lower and upper magnon bands. The ratio of $\kappa_{xy}$ with respect to $\kappa_{xx}$ gives the magnon Hall angle $\theta_{xy}$ that is shown correspondingly in Fig. 3(c). It decreases dramatically with temperature while at low temperature; it can even exceed 1, indicating that the magnon Hall current can be larger than the longitudinal magnon current.

Figure 4 shows some representative magnon noises as a function of temperature in the four-terminal case. Overall, the shot noise is considerable small compared to the thermal magnon noise, which is consistent with Eq. 12. They all increase monotonically with an increase of temperature and reach zero in the limit of $T \to 0$. While the nonlocal thermal noise $\mathcal{T}_{24}$ and the two shot noises $\mathcal{S}_{22}$, $\mathcal{S}_{24}$ are strongly suppressed in the presence of the Gilbert damping at $\gamma = 0.005$, the local thermal noise $\mathcal{T}_{22}$ surprisingly remains unchanged.

In summary, we have developed a non-equilibrium Green's function theory for the temperature gradient driven magnon transport in ferromagnetic insulators. This opens a new strategy to explore a variety of problems related to magnon transport. We provide explicit expressions for the magnon current, magnon noise, magnon Hall angle along with Fano factor, and uncover that the Fano factor is an important observable to gain deep insights into the magnon transport. We find that, in sharp contrast to that in a quantum anomalous Hall insulator, the magnon current and noises in TMIs can not be directly formulated. Despite the TMI considered in this Letter, our theory should also allow a quantitative treatment of magnon transport properties in other ferromagnetic insulators. In addition, our theory can also be readily generalized to the transport of other physical observables carried by magnon such as energy and spin angular momentum.

**Acknowledgments:**
This work is supported by the National Basic Research Program of China (Grants Nos. 2024YFA1409003 and 2022YFA1403700), National Natural Science Foundation of China (Grants Nos. 12147126 and 12404056). Yu-Hang Li also acknowledges the financial support from the Natural Science Foundation of Tianjin, China (Grant No. 24JCQNJC01910), the Fundamental Research Funds for the Central Universities (Grant No. 010-63263112).

**Statements and Declarations:** The authors declare no competing interests.

**Data availability:** The data that support the findings of this study are available from the corresponding author on reasonable request.

**Author contributions:** Y.-H.L and H.J. conceived the idea of topological magnon noises. S.L. and T.Z developed the formalism and performed the calculations with assistance from Y.-H.L. Y.-H.L. wrote the manuscript with contributions from all authors. Y.-H.L and H.J. supervised the project.